\documentclass[english,groupedaddress, superscriptaddress,aps,pre,10pt, nofootinbib, notitlepage]{revtex4-2}
\usepackage{hyperref}
\usepackage{geometry}
\usepackage{bigints}
\usepackage{graphicx}
\usepackage{gensymb}
\usepackage{mathtools}
\usepackage{amsmath}
\usepackage{amssymb}
\usepackage[normalem]{ulem}

\DeclarePairedDelimiter\abs{\lvert}{\rvert}%
\DeclarePairedDelimiter\norm{\lVert}{\rVert}%

\usepackage{color}
\definecolor{blue}{rgb}{0,0,1}
\definecolor{black}{rgb}{0,0,0}

\newcommand{\revise}[1]{\textcolor{black}{#1}}
\definecolor{dgreen}{rgb}{0,0.5,0}

\definecolor{dred}{rgb}{0.5,0,0}

\definecolor{dyellow}{rgb}{0.75,0.75,0}

\makeatletter
\let\oldabs\abs
\def\abs{\@ifstar{\oldabs}{\oldabs*}}
\let\oldnorm\norm
\def\norm{\@ifstar{\oldnorm}{\oldnorm*}}
\makeatother

\makeatletter
\usepackage{enumerate}
\usepackage[caption=false]{subfig}

\begin{document}
\title{Notes on the Superstatistical approach to UK Airport Arrival Delays Statistics}
\title{Superstatistical approach to UK Airport Arrival Delays Statistics}
\title{COVID-19 Impact on Plane Delays }
\title{Statistical characterization of Airport Arrival Delays}
\title{Statistical Characterization of Airplane Delays}

\author{Evangelos Mitsokapas}
%\thanks{contributed equally}
%\affiliation{School of Mathematical Sciences, Queen Mary University of London, London E1 4NS, United Kingdom}
\affiliation{School of Mathematical Sciences, Queen Mary University of London, London E1 4NS, United Kingdom\\Correspondence to b.schaefer@qmul.ac.uk}

\author{Benjamin Schäfer}
%\thanks{contributed equally}
%\email{Correspondence to b.schaefer@qmul.ac.uk}
\affiliation{School of Mathematical Sciences, Queen Mary University of London, London E1 4NS, United Kingdom\\Correspondence to b.schaefer@qmul.ac.uk}

\author{Rosemary J. Harris}
\affiliation{School of Mathematical Sciences, Queen Mary University of London, London E1 4NS, United Kingdom\\Correspondence to b.schaefer@qmul.ac.uk}
\author{Christian Beck}
\affiliation{School of Mathematical Sciences, Queen Mary University of London, London E1 4NS, United Kingdom\\Correspondence to b.schaefer@qmul.ac.uk}

\begin{abstract}

The aviation industry is of great importance for a globally connected economy. Customer satisfaction with airlines and airport performance is considerably influenced by how much flights are delayed. But how should the delay be quantified with thousands of flights for each airport and airline? Here, we present a statistical analysis of arrival delays at several UK airports between 2018 and 2020. We establish a procedure to compare both mean delay and extreme events among airlines and airports, identifying a power-law decay of large delays. Furthermore, we note drastic changes in plane delay statistics during the COVID-19 pandemic. Finally, we find that delays are described by a superposition of simple distributions, leading to a superstatistics.
\end{abstract}

\maketitle
\makeatother

\section{Introduction}

%air transportation important
The aviation industry was a rapidly growing sector until recently, prior to the current COVID-19 pandemic. Economic growth led to higher average yearly distances travelled, as well as higher air traffic volumes, robustly observed among several regions worldwide until 2019 \cite{hakim2016causal,brida2018exploring}. But both the ongoing pandemic \cite{suau2020early} and also the push towards more renewable options in aviation \cite{kuhn2011renewable} may induce a considerable change in the industry in the future. This makes the industry a very interesting object to study as it transforms. 

%Introduction to delays
As a passenger, an important benchmark for evaluating travel options, e.g.\ in terms of airports, airlines or even modes of transportation (train vs plane) is the punctuality of each option. In particular, flight delays severely decrease customer satisfaction \cite{efthymiou2019impact} and might lead to customers choosing a different airport or airline, \revise{in the long term.} Generally, it is important to quantitatively understand delay-risks both in terms of the expectation values but also in terms of the extreme events, i.e.\ quantifying how likely a very early or very late arrival is.

%Previous work on delays in airplanes
The study of delays in aviation is already an active field of research. Previous, simple, investigation frameworks to classify and categorize delays have been proposed \cite{rebollo2014characterization} but mostly rely on mean values. In other cases, stochastic models of plane delays \cite{rosenberger2002stochastic} were developed either  without considering the corresponding probability distributions or assuming simple Normal or Poisson distributions \cite{mueller2002analysis}.
More recent work also includes the application of machine learning techniques to aviation data, e.g.\ via recurrent neural networks \cite{gui2019flight}.
One problem of any data-driven approach is that many articles on aviation research solely rely on  proprietary data: In a recent review investigating 200 research articles, $68\%$ were based on proprietary data \cite{li2019reviewing}. Hence, to enable the broader applicability of machine learning applications, more publicly available data are still required.

%Previous work on transportation delays and superstatistics
To quantify delay statistics, we will go beyond the often-used averages of delays \cite{rebollo2014characterization} and instead investigate the entire probability density function \revise{of delays at a given airport. Thereby, we consider all possible delay values, from highly negative delays (i.e.\ flights arriving significantly earlier than  their scheduled arrival time) to severely positively delayed flights. These delay distributions are influenced by many different aspects, including random events, congestion,
delay propagation between airports \cite{fleurquin2013systemic, pyrgiotis2013modelling} and (for long-haul flights on large scales) the topological structure of the worldwide air transportation network \cite{new1,new2}.}
To explain the emergence of heavy tails in a local distribution, i.e.\ extreme deviations from the mean, we will utilize superstatistical modelling \cite{BCS}. Such an approach has been successfully applied in transport before, for modelling train delays \cite{briggs2007modelling}; it has also attracted recent interest when describing fluctuations in the energy system  \cite{schafer2018non} and air pollutant concentrations \cite{williams2020superstatistical} and it has been extended to the general framework of diffusing diffusivities in nonequilibrium statistical physics and biologically inspired physics \cite{metzler2020superstatistics, chubynsky2014diffusing,itto2021}.

%Overview of our article
In this article, we present new data collected from 2018 to 2020 at several UK airports, with a particular focus on Heathrow, being the most important international hub in the UK. 
The data were publicly available from the arrival information of each airport, given out on their websites each day but had to be collected and processed for further usage. \revise{While the past arrival data can no longer be accessed via the airport websites, all collected data have been uploaded in a repository, see Methods.}  We analyse the full probability density of delay distributions and introduce certain performance indices to describe these distributions, such as the mean delay, the exponential decay rate of negative delays, and the power-law exponent of large positive delays. These indices are then compared for the different UK airports and the different airlines operating at these airports, to understand the main features of the delay statistics \revise{(such as frequency of extreme delays, average delay per airport or per airline, etc)} in a more systematic way. Finally, we deal with a theoretical model to explain features of the delay statistics. We show that the power law of large positive delays can be linked to a superposition of exponential delays with a varying decay parameter, in a superstatistical approach. Conversely, negative delays (early arrivals) do not exhibit any power laws but simply behave in an exponential way, with extremely early arrivals exponentially unlikely. Throughout this article, we assume that passengers prefer to arrive as early as possible, i.e.\ with as little positive and as much negative delay as possible.

\section{New data}
\label{sec:data} 

%General data introduction
We collected flight details from a number of  different airports. For the purposes of this article, we have taken into consideration the top five UK airports, in order of passenger traffic \cite{CAA}, namely: London Heathrow Airport (LHR), London Gatwick Airport (LGW), London Luton Airport (LTN), London Stansted Airport (STN) and Manchester Airport (MAN).  
For a period of time lasting between Autumn 2018 and Spring 2019, we collected  a combined total of approximately two-hundred and twenty thousand ($2.2 \times 10^5$) flight-arrivals from all five airports mentioned above. Furthermore, we continued collecting flight-information from London Heathrow during the 2020 COVID-19 pandemic, to illustrate the effect the lockdown had on the delay distribution.
For each flight, we recorded the airline company operating the flight along with the corresponding flight number, departure and arrival airports, as well as scheduled and actual landing times.
The delay is then computed simply as the difference between an aircraft's scheduled arrival time and its actual arrival time. \revise{Note that 
airlines and airports presumably have some freedom in setting the scheduled arrival time, potentially influencing the average ``delay'' (average difference between scheduled and actual arrival).} We made all collected data publicly available. 
For details of the data processing and availability, see Methods. 

%Review of Heathrow
The main body of our data (about $85\%$) is sourced from London Heathrow, making it the chief focus of our analysis simply due to its size. London Heathrow is an international airport operating flights of 80 different airlines in total, which fly to 84 different countries around the world, as of 2019 \cite{CAA}. Of course, in addition there are domestic flights within the UK. The passenger nationalities are $48\%$ European and UK and $52 \%$ from the rest of the world. It is the busiest airport in Europe by passenger traffic \cite{CAA}.

\begin{figure}[!ht]
    \centering
    {{\includegraphics[width=0.9\textwidth]{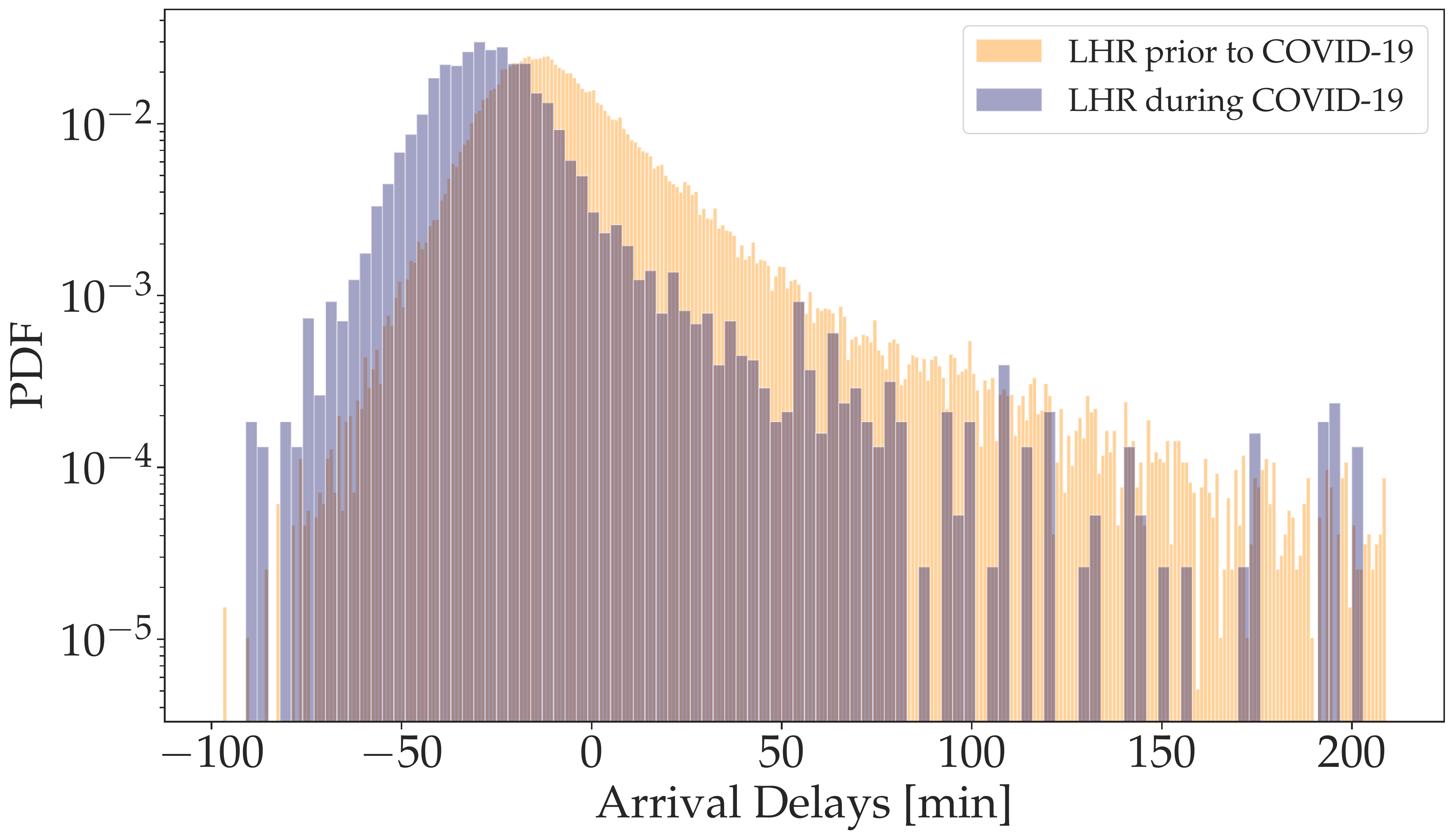} }}%
    \caption{Flight delays follow a broad distribution with large negative and positive delays. We display LHR delay histograms prior to and during the COVID-19 pandemic, both normalized. As the COVID-19 LHR data set is significantly smaller in size, compared to the regular LHR data set, it contains many gaps, where no data were recorded. The COVID-19 data set is significantly shifted towards the left (smaller delays) as compared to the pre-pandemic time.}
    \label{fig:Heathrow_full_hist}
\end{figure}

%Heathrow data initial analysis
The empirical probability density function (PDF) of all delays is a key characteristic to monitor, see Fig.~\ref{fig:Heathrow_full_hist} for all Heathrow delays. There, we compare the data collected from 2018 to 2019 with more recent data collected during the 2020 COVID-19 pandemic (during the first lockdown in Spring to Summer 2020),
which led to a drastic reduction in air transport \cite{GuardianHeathrowPassengerReduction2020,nivzetic2020impact}. There are two interesting observations: Firstly, the delay statistics under COVID-19 are shifted to the left, indicating overall smaller delays (including more negative delays); secondly, the general shape of the distribution does not change drastically. In particular, we observe a fast decay of the PDF of negative delays on the left side and a much slower decay of the PDF on the right side for positive delays. 
In the following sections, we will
analyse this behaviour in much more detail.

%A key question now is: How do we quantify the change in the probability density?

\section{Quantifying delay statistics}
\label{sec:quantify}

%Section overview
Starting from a histogram of the flight delays, we derive three indices/measures to quantify flight delay distributions: Mean delay, exponent of left exponential and power-law exponent of right $q$-exponential, as explained below in detail. We will use the LHR data previous to any COVID-19 influence as our main example.

\begin{figure}
    \centering
    \begin{minipage}{0.45\textwidth}
        \centering
        \includegraphics[width=1.0\textwidth]{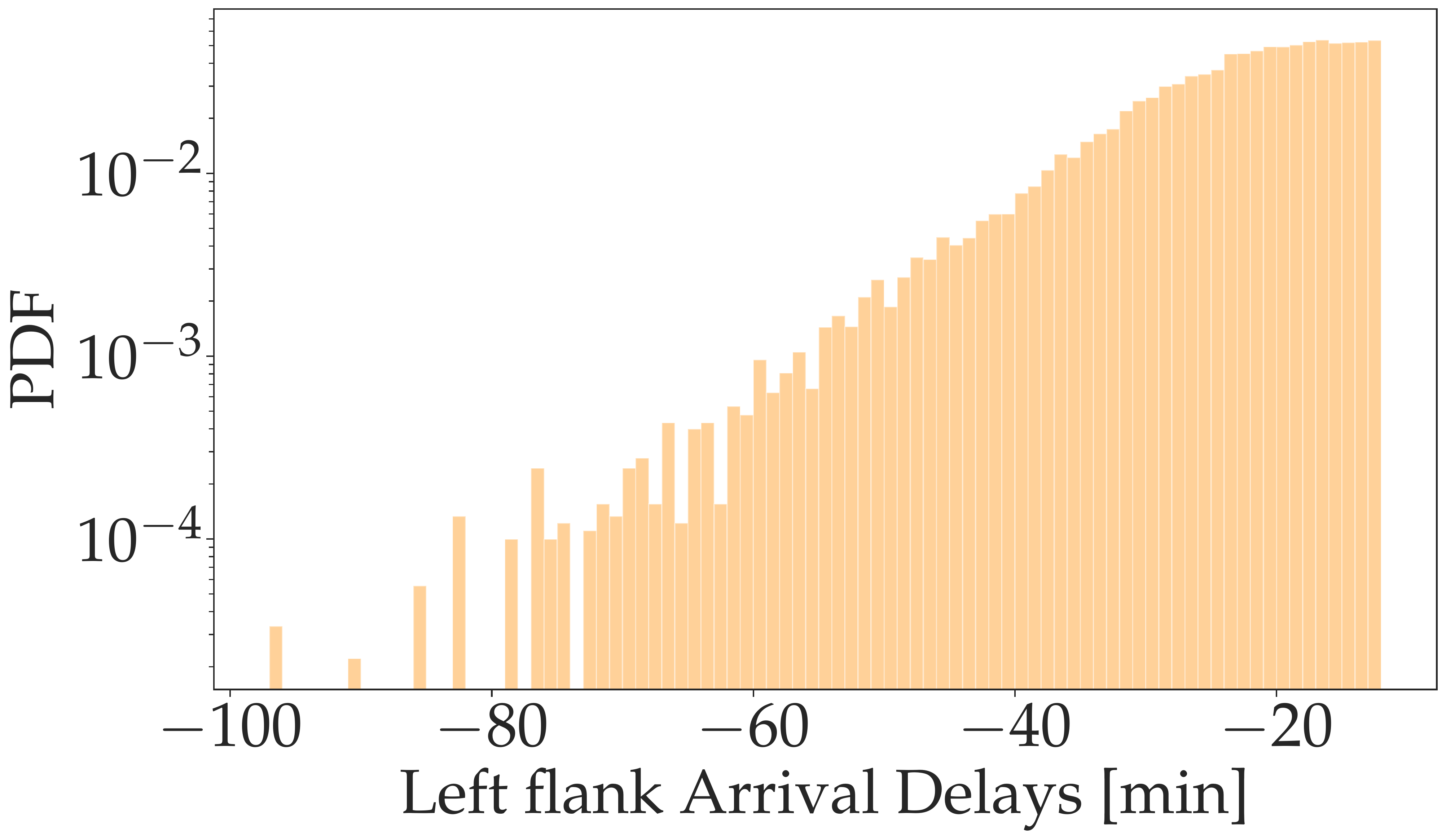} 
    \end{minipage}
    \vspace{0cm}
    \begin{minipage}{0.45\textwidth}
        \centering
        \includegraphics[width=1.0\textwidth]{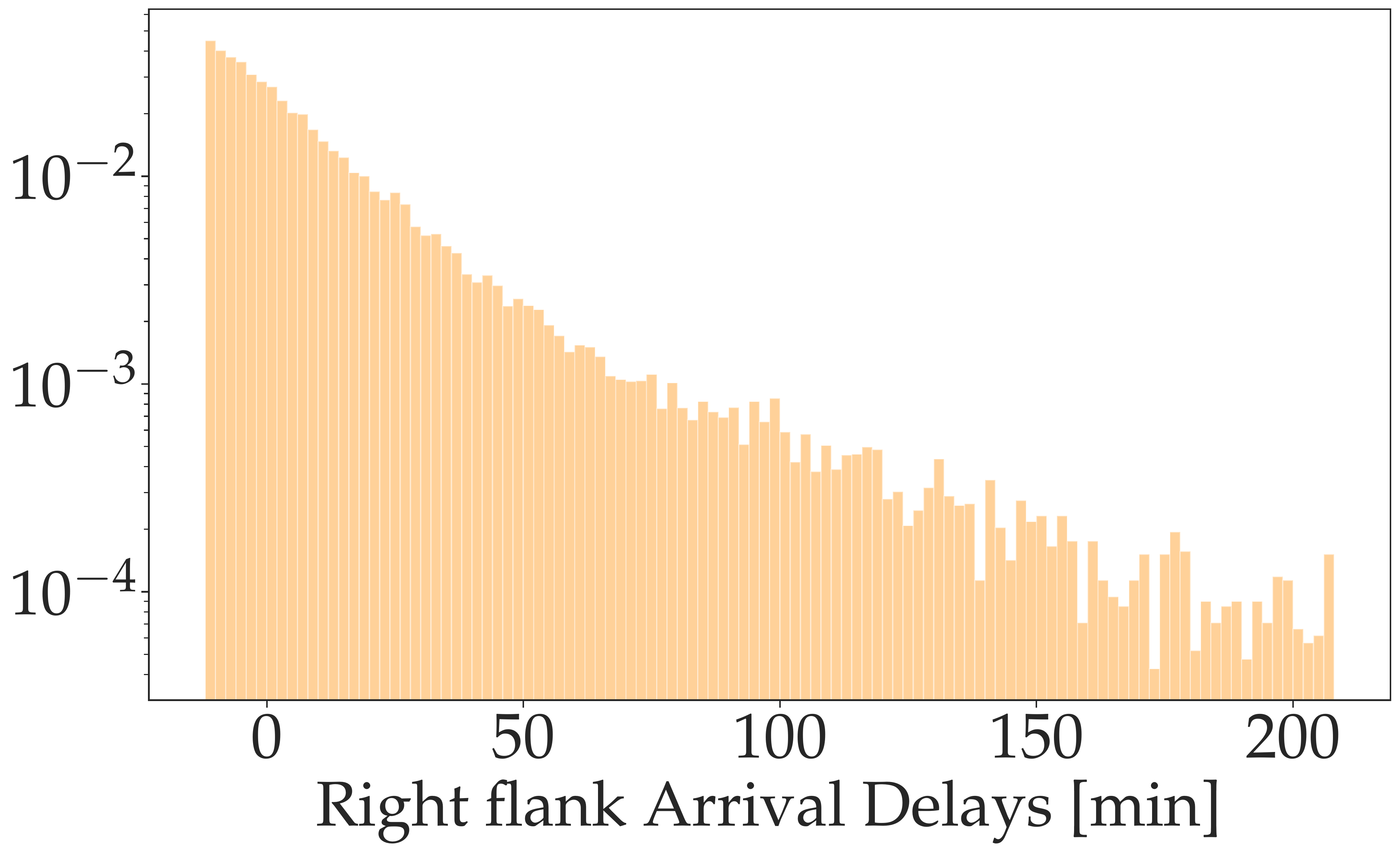}
    \end{minipage}
    \caption{Splitting the full distribution at the peak leads to two easier-to-fit flanks. Left: Negative delays decay approximately linearly in the log-scale and thereby suggest an exponential fit \eqref{expdist}. Right: Positive delays display substantial heavy tails and thereby suggest the usage of a $q$-exponential function \eqref{qexpo}.}
    \label{fig:Splitting_the_histogram}
\end{figure}

%splitting histograms into left and right flank
As a first step, we split the full histogram at its peak value into two histograms, a left flank of predominantly negative delays and a right flank of predominantly positive delays, see Fig.~\ref{fig:Splitting_the_histogram}.
Based on the shape of the empirical distributions, we use exponentials and  $q$-exponentials as fitting functions, see also Methods for details. 
\revise{Splitting the histogram has two advantages: Firstly, the analysis of each flank is much simpler than the analysis of the full aggregated data. Secondly, a given stakeholder might be particularly interested in positive rather than negative delays, or vice versa.}

%Left flank: exponential
The left flank is observed to be well approximated by an exponential function of the form \begin{equation}
\label{expdist}
p(t_L; \lambda) = \lambda e^{- \lambda t_L}, \lambda >0,
\end{equation}
where $t_L$ are the rescaled arrival delays on the left flank, see Methods for details.
The exponent $\lambda$ here quantifies the exponential decay of the probability
of early arrivals.
Therefore, a large $\lambda$ implies that few flights arrive very early while a small $\lambda$ indicates that very large negative delays are observed. \revise{Since we assume that passengers prefer to arrive as early as possible, a small $\lambda$ indicates good performance.}

%right flank: q-exponential 
The right flank of the delay distribution  obeys a power law, i.e.\ a slow decay of $p\sim t^\nu$, with $\nu$ negative.
To quantitatively describe the right flank, we use a $q$-exponential function \cite{tsallis1988possible} of the form \begin{equation}
\label{qexpo}
 p (t_R; q, \lambda_q) = (2-q)\lambda_q \left[ 1+ (q-1) \lambda_q t_R \right]^{\frac{1}{1-q}},
\end{equation}
where $t_R$ are the rescaled arrival delays on the right flank, see Methods for details.
The power-law exponent, i.e. the rate at which the probability density
decays for high (positive) delay values, is given by $\nu:= 1/(1-q), 1<q<2$.
Note that the scale parameter $\lambda_q>0$ is relevant for the precise fit but \revise{does not impact the power-law exponent $\nu$. Since the power-law decay is controlled by the value $q$, we utilize $q$ to characterize the right flank.} \revise{ Contrary to the left-flank exponential decay, good performance is indicated by the absolute value of the right-flank power law exponent $\nu$ being large. The reason is that large (absolute) values of $\nu$ imply a rapid decay of the probability density of positive delays, i.e.\ fewer extreme events of very delayed arrivals.}

\begin{figure}
    \centering
   {{\includegraphics[width=0.9\textwidth]{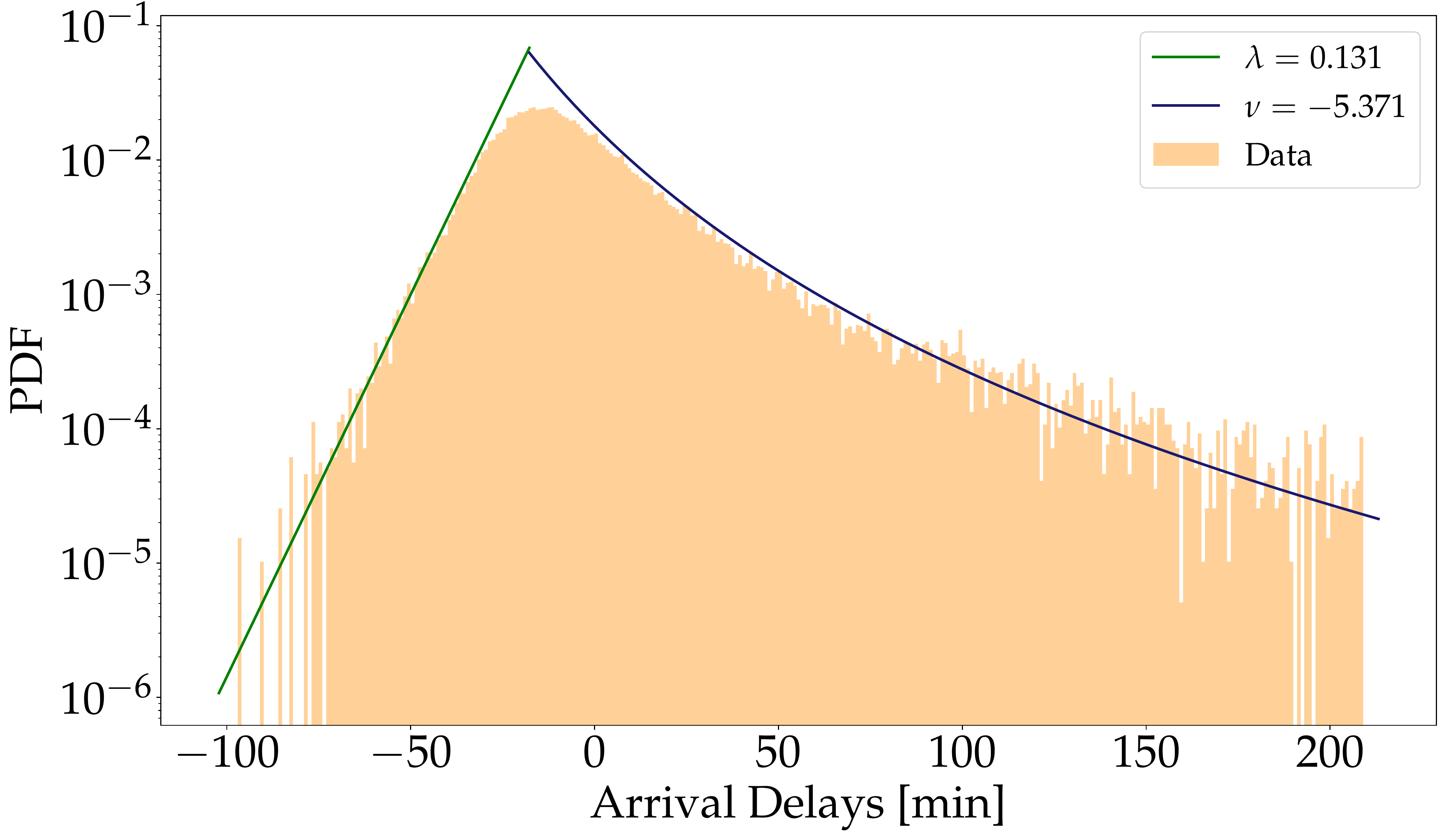} }}%
    \caption{Exponential (green) and $q$-exponential (blue) theoretical distributions capture the empirical distribution. The fits are obtained via the MLE method, see Methods for fitting details. To complement the over-estimated ``peak'' (tent-like shape) we introduce the mean delay $\mu$ index.}
    \label{fig:Combining_fits}
\end{figure}

%introduce the mean as third necessary measure
Finally, we note that the two flanks describe the tails of the distribution well, but overestimate the height of the peak, i.e.\ the most likely value, see Fig.~\ref{fig:Combining_fits}.
\revise{To include more information on the most frequent delays,} we complement the two previous fits by using the mean delay $\mu$ as a third index. \revise{Here we interpret a small positive $\mu$, or a negative $\mu$ (indicating early arrival), as desirable for passengers.}  In the case of LHR, the three delay indices that we introduced are $\lambda = 0.131$, $\mu = -5.06$ and $\nu = -5.371$. We also introduce a continuous smooth fitting function for the full range in the "Connecting the flanks" section.

%Brief discussion of parameters being gamed/manipulated by the operators
Note that the mean value $\mu$ can be easily
manipulated by airline companies
by scheduling flight arrival times later then actually needed, hence always causing a negative mean delay, which may artificially improve their performance. On the contrary, the tail behavior truthfully represents the extreme event statistics for both positive and negative delays and cannot be easily manipulated by the operators. 

\section{Comparison of airports and airlines}
\label{sec:comparison}
We here use the previously developed framework to quantify and compare delay statistics for different airlines and airports. Intuitively, we expect that long-distance flights would, on average, yield more extreme early or late arrivals, compared to the corresponding short-distance ones. 
Thus, we distinguish between short-distance airlines, covering mostly domestic and European destinations, and airlines that include 
long-distance, international destinations, as well as destinations within Europe.  
We first compute the three indices $\lambda , \mu , \nu$ for each of those airline groups and then compare full airport statistics, aggregating all airlines. 

\begin{figure}
    \centering
    {{\includegraphics[width=\textwidth]{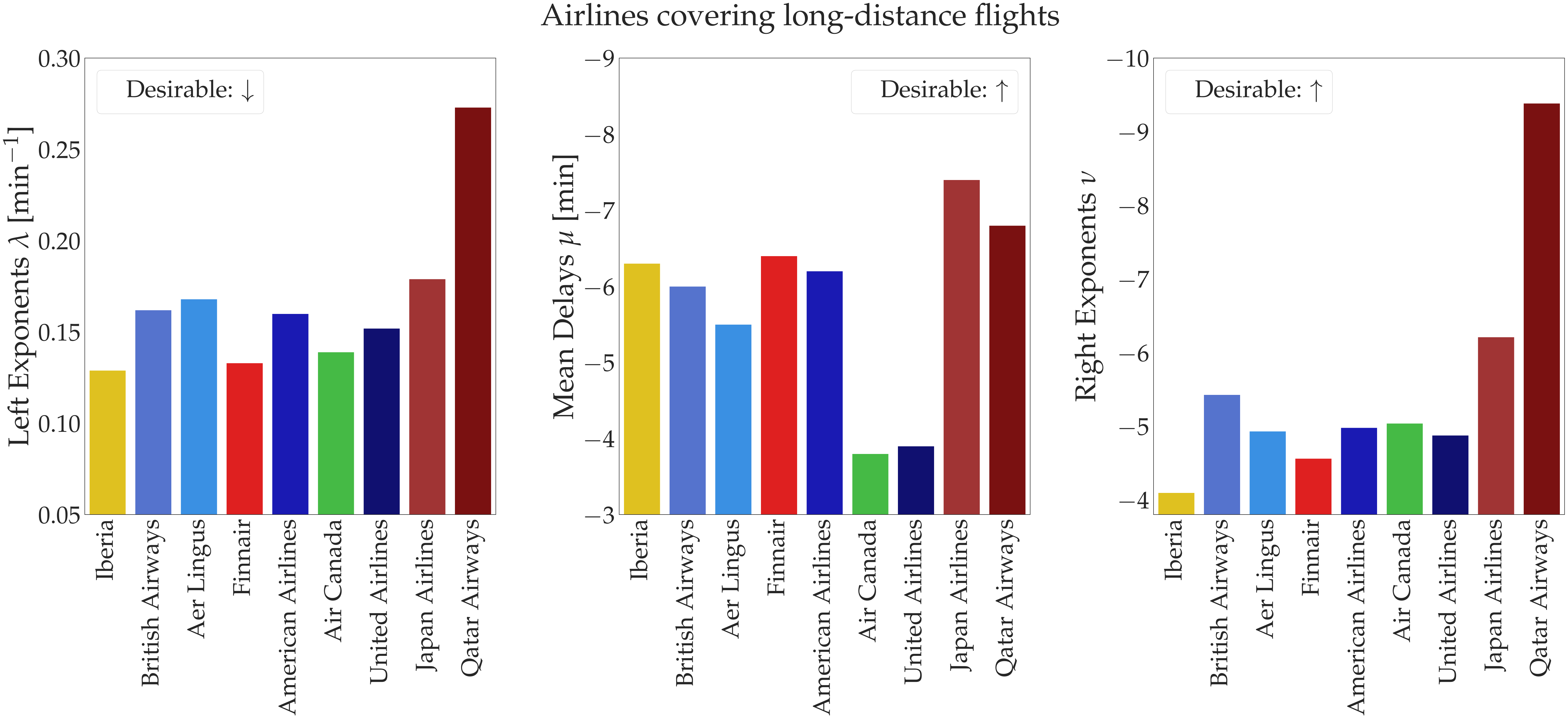} }}%
    \caption{International airlines appear to differ substantially in their three delay indices. We plot the left-side (negative) delay exponential decay, right-side (positive) delay power-law decay and the mean delay. Arrows indicate whether a small or large value is desirable.}
    \label{fig:Compare_AirlinesInternational}
\end{figure}

% discuss differences/similarities between international flights
There are several factors impacting the delay distribution for each airport or airline: Airline policies, flight routes, technical defects or issues with documentation contribute to $27\%$ of all delays \cite{EUROCONTROL}.
Specifically, overseas flights are more sensitive to wind (head wind or tail wind), as well as unstable weather conditions (storms, fog) and military exercises. Airlines operating international flights, as illustrated in Fig.~\ref{fig:Compare_AirlinesInternational},  exhibit considerable variations in
their flight delay indices. 
Note that a low left exponent $\lambda$ may be regarded as a desirable property (flights often arrive very early) while good performance is definitely indicated by low mean $\mu$ and right exponent $\nu$ (low mean delay and few very late arrivals). Since the latter two quantities tend to be negative, their absolute values should be large.
Comparing the airlines, we observe a ``grouping'' behaviour for some of the carriers. On the one hand, airlines having a blend between short-distance (e.g.\ domestic or EU) and overseas destinations, such as Iberia, British Airways (BA), Aer Lingus and Finnair, appear to follow a similar trend for each index.  On the other hand, airlines that do not possess such a spread of destinations tend to perform well only in some of the indices. As an illustrative example, we choose Air Canada and United Airlines: Although both their left and right exponents are in a similar range to the other airlines, their mean delays are substantially less negative than those of their competitors. 

\begin{figure}
    \centering
    {{\includegraphics[width=15cm]{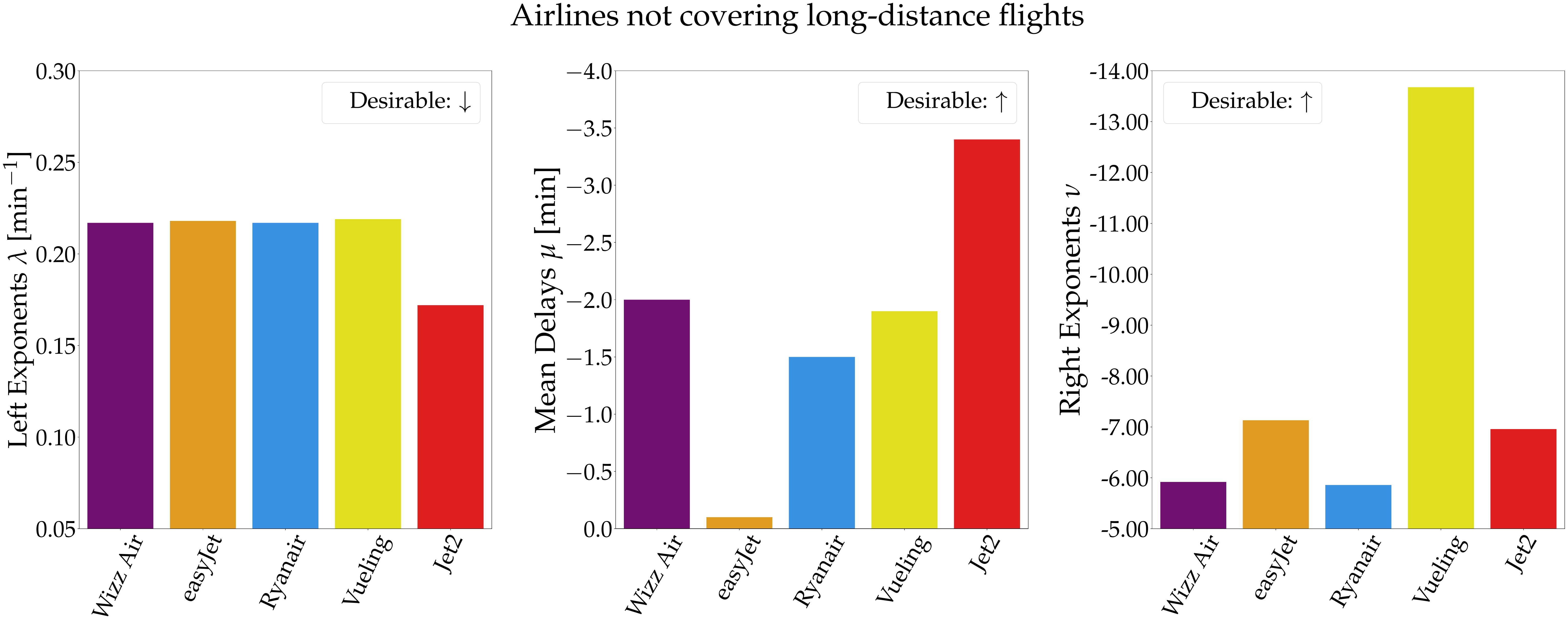} }}%
    \caption{Delay indices for low-cost airlines not covering long-distance flights. Wizz Air, easyjet, Ryanair and Vueling share the largest $\lambda$ index (early arrivals). Jet2 has the lowest mean delay $\mu$ and Vueling is characterized by the lowest $\nu$ index (late arrivals).}
    \label{fig:Compare_AirlinesLow}
\end{figure}

% discuss differences/similarities between short-distance flights
Characterization of short-distance flights shows a strong grouping of the delay behavior for some airlines. As seen in Fig.~\ref{fig:Compare_AirlinesLow}, comparison of five of the largest low-cost domestic and European providers, reveals a systematic similarity between Wizz Air, easyJet and Ryanair. All three airlines manage to perform well in the left exponent metric, maximizing early arrivals, while they maintain an acceptable negative average delay (with easyJet obtaining the lowest value here). Again, they are characterized by similar right-exponents, translating to a certain share of overall late arrivals. 
Furthermore, Jet2 outperforms all other short-distance airlines in $\lambda$ left-exponents and mean delays. Finally, Vueling resembles Wizz Air and Ryanair values in the $\lambda$ and $\mu$ metrics but seems to have less late arrivals as per its high right exponent $\nu$.

Comparing the long distance airlines with the short-distance ones, we notice some differences: Airlines covering long distances tend to display lower (more desirable) left exponents as well as more negative mean delays. Meanwhile, the right exponent behavior is similar between the two groups with Vueling and Qatar Airlines as the ``outliers'' in their respective categories. Whether this behavior is due to company policies or flight distance remains a question for future research.

\begin{figure}
    \centering
    {{\includegraphics[width=15cm]{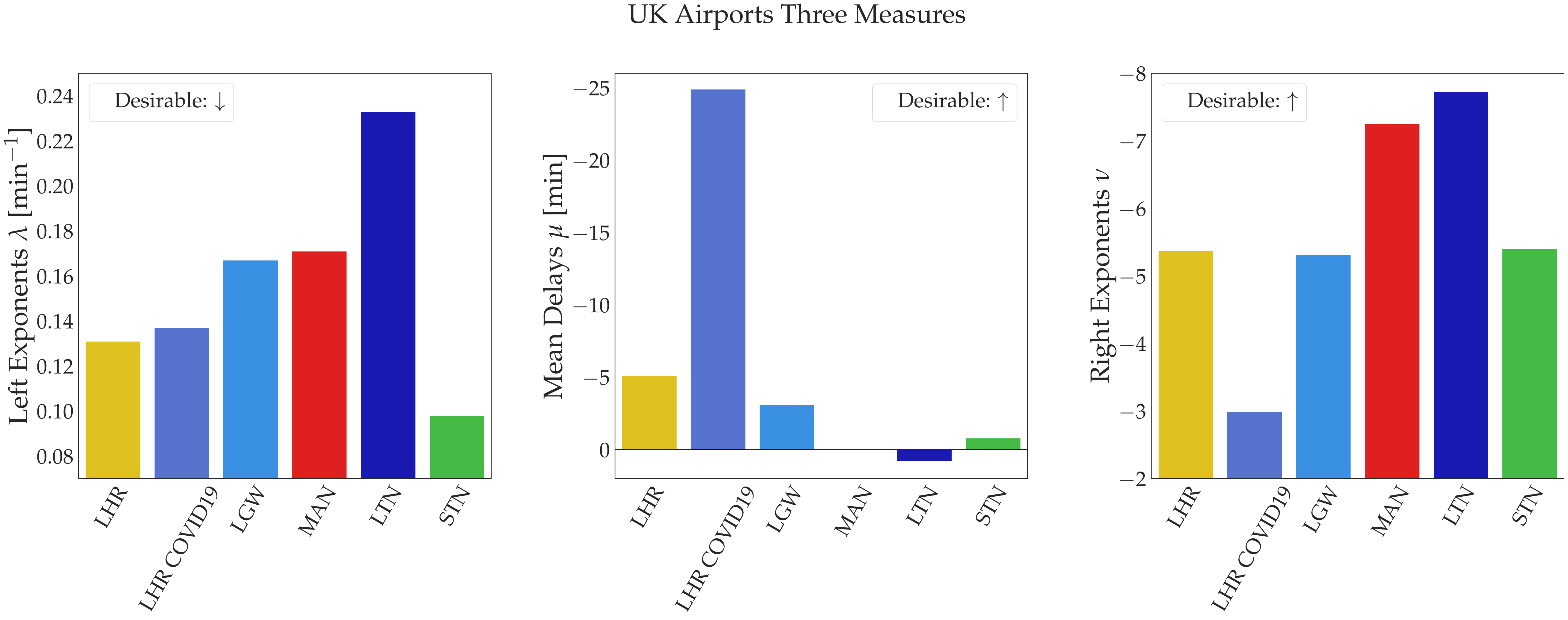} }}%
    \caption{Airports appear to differ substantially in the three delay metrics. Airports that serve mostly domestic and European destinations, such as LTN and STN, behave differently from international airports such as LHR, LGW and MAN.}
    \label{fig:Compare_Airports}
\end{figure}

Studying the indices for individual airports yields interesting insights as well. Airports populated by airlines flying mainly to domestic and EU destinations, such as LTN and STN, \revise{have a mixed score  in both early and late arrivals}, with an approximately net zero mean delay, see Fig.~\ref{fig:Compare_Airports}. On the one hand, STN is characterized by the minimum $\lambda$ value, showing the best performance in early arrivals in the group of airports, while LTN attains \revise{the maximum} value. On the other hand, it can be seen that LTN scores the best $\nu$ value while STN lies very slightly above the group median $\nu$. 
Interestingly, mean delays at MAN airport are net \revise{zero}, contrary to LHR and LGW where arrivals are scheduled in such a way that the mean delay is negative. Furthermore, MAN seems to \revise{have a similar performance to LGW} in the early arrivals index, having \revise{a slightly worse} score, but does \revise{attain the second best value} when compared from the perspective of extreme positive delays.  International airports LHR and LGW (with the exception of LHR COVID-19) tend to cluster around similar values for all delay indices.

LHR during the COVID-19 pandemic outperforms all airports on the mean delay index by a large margin. \revise{Indeed focussing in on LHR, we see a clear difference between the time prior to the pandemic ($\mu_\text{LHR}\approx -5$min) and during the pandemic ($\mu_\text{LHR COVID19}\approx -25$min).} The reason behind this is that the dramatic reduction of flight traffic worldwide saw many flights arriving too early.
Interestingly, the left exponent, i.e.\ the decay of early arrivals, did not change substantially, compared to LHR under business-as-usual conditions since the shape of the delay distribution on the left did not change much but was only shifted to more negative values.
The right flank behaves quite differently: Both business-as-usual and LHR during the COVID-19 pandemic, recorded relatively heavily delayed flights, which arrived more than 3 hours late (see also Fig.~\ref{fig:Heathrow_full_hist}). The right index reveals the likelihood of these extreme events. In the case of LHR under COVID-19, the low mean delay suggests early arrival but relative extreme events are still present and hence the right exponent reveals this poor performance.

\revise{Notice that  we cannot fully exclude a sampling bias of the airline analysis due to the different number of flights recorded for each airport: For a given airline, e.g.\ BA, we use all flights at all airports in our data set. However, since we recorded more total flights in LHR, the BA distribution is influenced more by the LHR data than by other airports.}

\section{Superstatistical modelling of delays}
\label{sec:superstatistics}

%Introduction to superstatistics
As we have seen previously, the right flank of the delay statistics exhibits heavy tails and is well-described by a $q$-exponential. Let us now explore a potential explanation for this particular distribution by employing the framework of superstatistics \cite{beck2001,beck-cohen,BCS}. Superstatistics is relevant when an aggregated system (e.g.\ a long time series) displays heavy tails, but the system may then be disentangled into many smaller sub-parts (e.g.\ short time periods of the trajectory). These sub-parts then are no longer heavy-tailed but follow a simple local distribution,
for example an exponential or a Gaussian.  This idea has been successfully applied, for example, to train delays \cite{briggs2007modelling}, electric power systems \cite{schafer2018non} and intermittent wind statistics \cite{weber2019wind}.

\begin{figure}
    \centering
    {{\includegraphics[width = 0.6\textwidth]{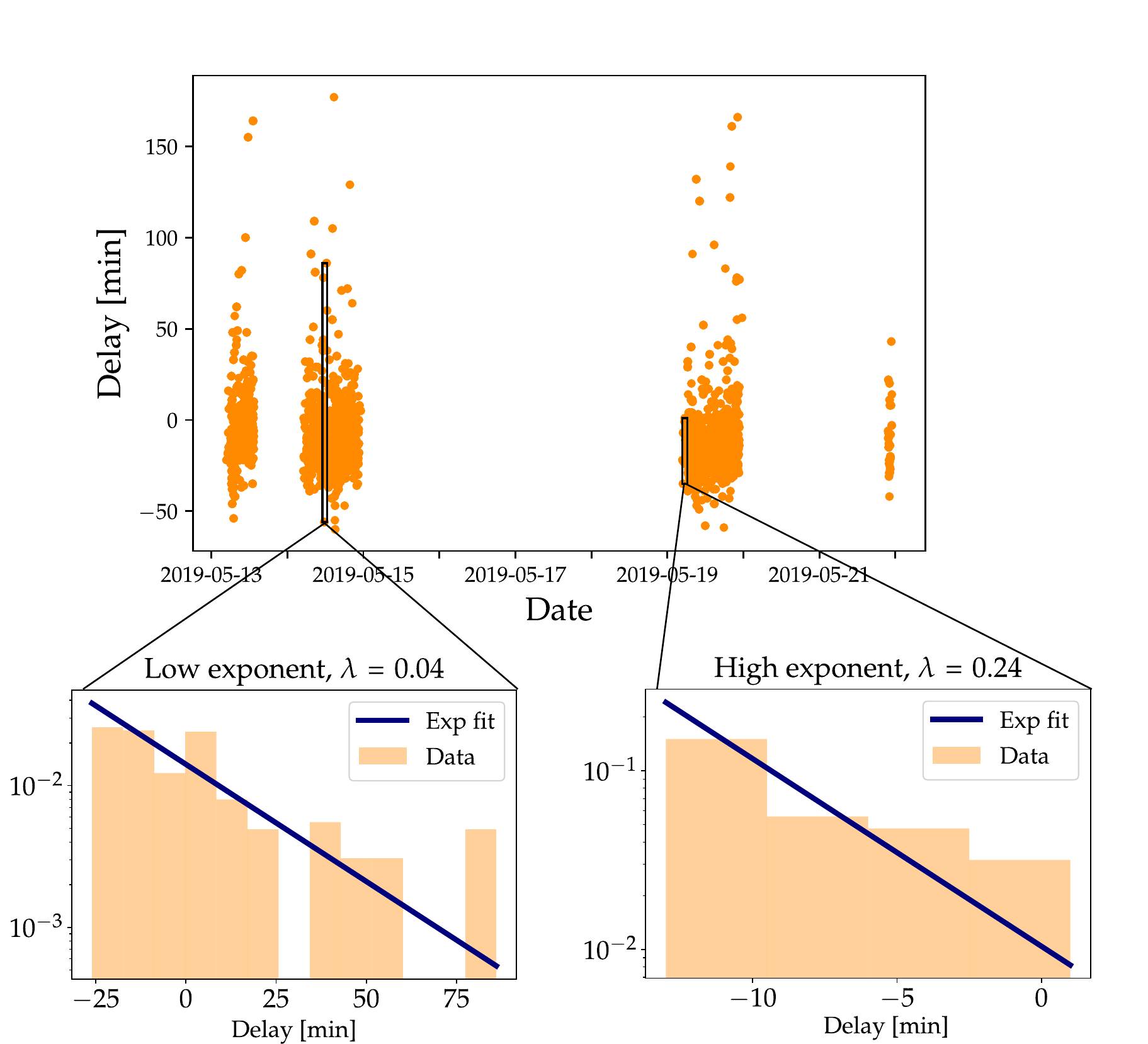} }}%
    \caption{We analyse the full time series of plane delays and extract a time window during which we observe locally exponential distributions. These local distributions can decay slowly or fast, i.e.\ the rate $\lambda$ is fluctuating.}
    \label{fig:Super_Snapshots}
\end{figure}

%Local exponential distributions
Assuming for now that the right-flank delays are indeed $q$-exponentially distributed and follow a superstatistics, we should be able to observe ``local" exponential densities, with a decay parameter $\lambda$. Superimposing all these $\lambda$, we get a $q$-exponential if the $\lambda$ themselves follow a $\chi^2$-distribution:
\begin{equation}
f(\lambda) = \frac{1}{\Gamma\left(\frac{n}{2}\right)}\left( \frac{n}{2 \lambda_0} \right)^{\frac{n}{2}} \lambda^{\frac{n}{2}-1}e^{-\frac{n\lambda}{2\lambda_0}}.
\end{equation}
Here $n$ denotes the number of degrees of freedom characterizing the fluctuations
in $\lambda$ and $\lambda_0$ is the sample mean of $\lambda$.
Indeed, choosing an appropriate time scale to separate the trajectory (see next paragraph), the heavy tails of the delay distributions vanish and instead the distributions are well described by simple exponential functions, see Fig.~\ref{fig:Super_Snapshots}.

\begin{figure}
    \centering
    {{\includegraphics[width = 0.6\textwidth]{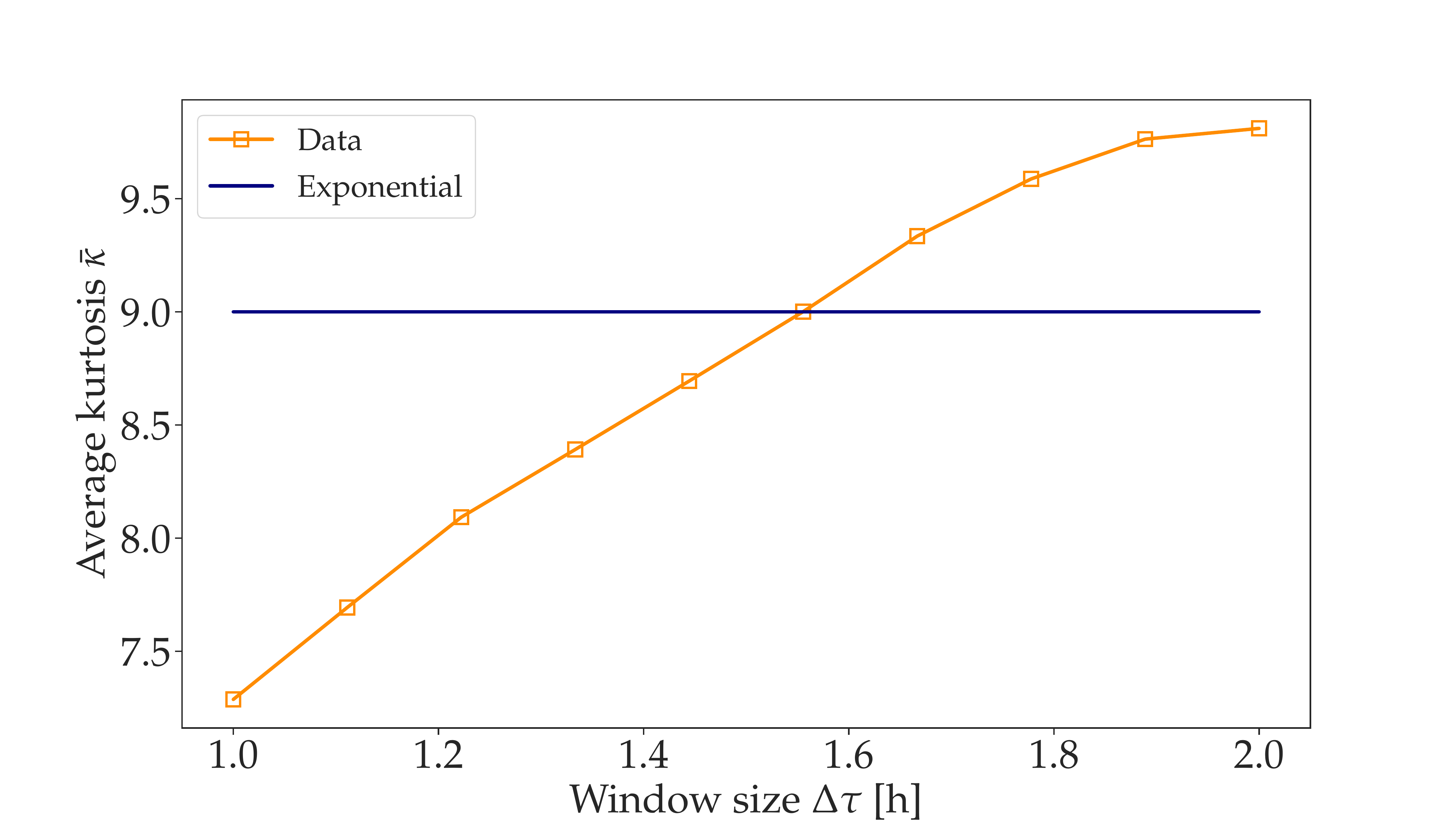} }}%
    \caption{The average kurtosis $\bar{\kappa}$ of the data set is plotted as a function of the time window $\Delta \tau$ in hours (yellow). The intersection between the
horizontal line at $\bar{\kappa} = 9$ (the kurtosis of an exponential distribution) and the $\bar{\kappa}$ vs $\Delta t$ curve gives the optimal value for $\Delta t$; we find $T \approx 1.55$ hours.}
    \label{fig:Super_LongTimeScale}
\end{figure}

%Long time scale 
Let us explain how to extract the
relevant time scale $T$ on which we locally observe exponential distributions. Since we know that an exponential distribution has a kurtosis of $\kappa_\text{exponential}=9$, we test time windows of different size $\Delta \tau$ and compute the local average kurtosis \cite{BCS} as 
\begin{equation} \label{Eq4}
\bar{\kappa}\left(\Delta \tau \right)=\frac{1}{\tau_{\text{max}} - \Delta \tau}\int_{0}^{\tau_{\text{max}}-\Delta \tau} d\tau_{0}\frac{\langle\left(u-\bar{u}\right)^{4}\rangle_{\tau_{0},\Delta \tau}}{\langle\left(u-\bar{u}\right)^{2}\rangle_{\tau_{0},\Delta \tau}^{2}},
\end{equation}
where  $\tau_{\text{max}}$ is the length of the time series $u$ and $\bar{u}$ is the mean of the time series. We denote by $\langle\dots\rangle_{\tau_0, \Delta \tau}$ the expectation formed for a time slice of length $\Delta \tau$ starting at $\tau_0$.
For the LHR data, \revise{we compute the local kurtosis and thereby determine the long time scale: $\bar{\kappa}\left( T \right)=9$, for $T\approx 1.55h$}, see Fig.~\ref{fig:Super_LongTimeScale}. 

\begin{figure}
    \centering
    \includegraphics[width=6cm]{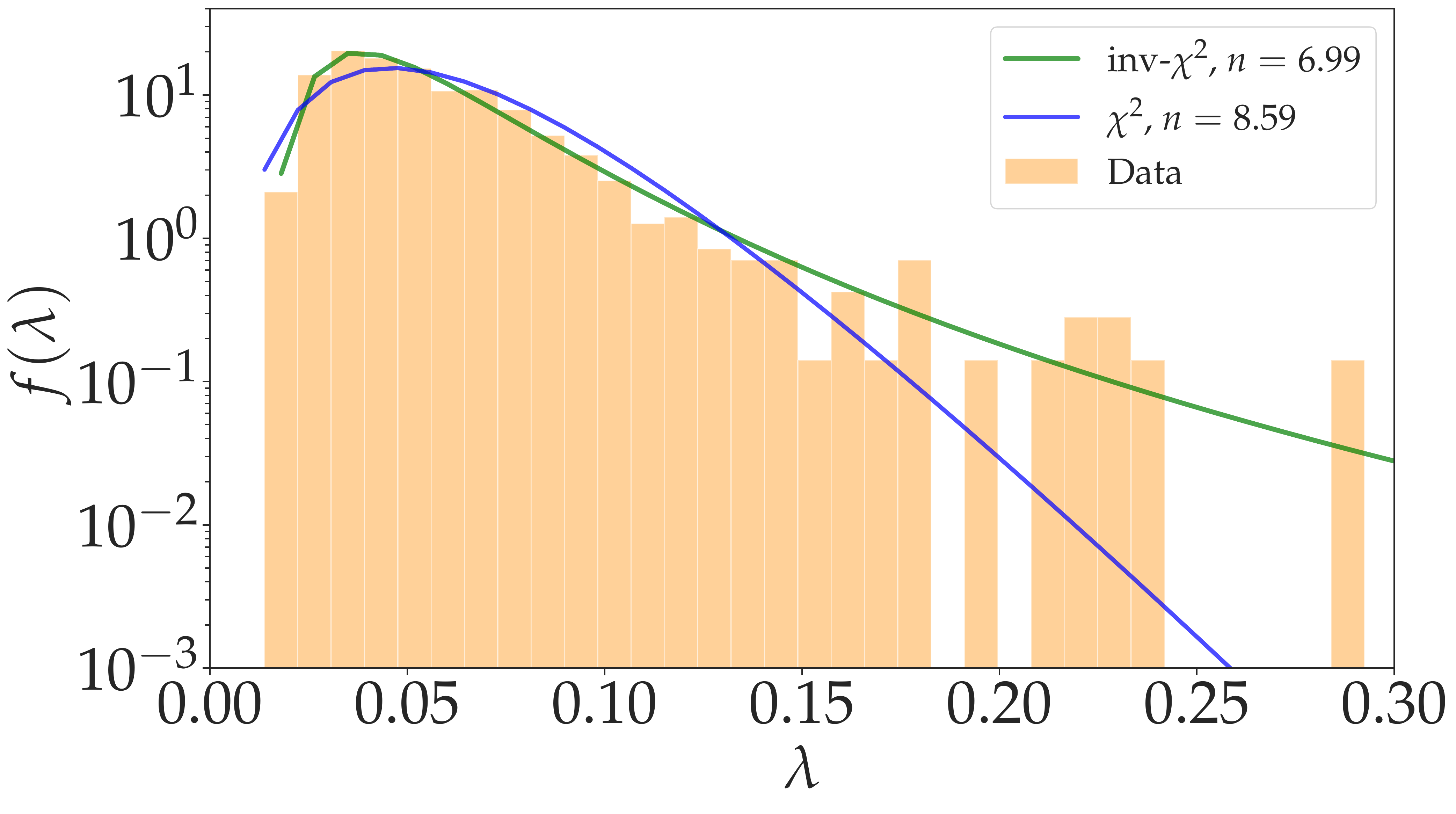}
    \includegraphics[width=6cm]{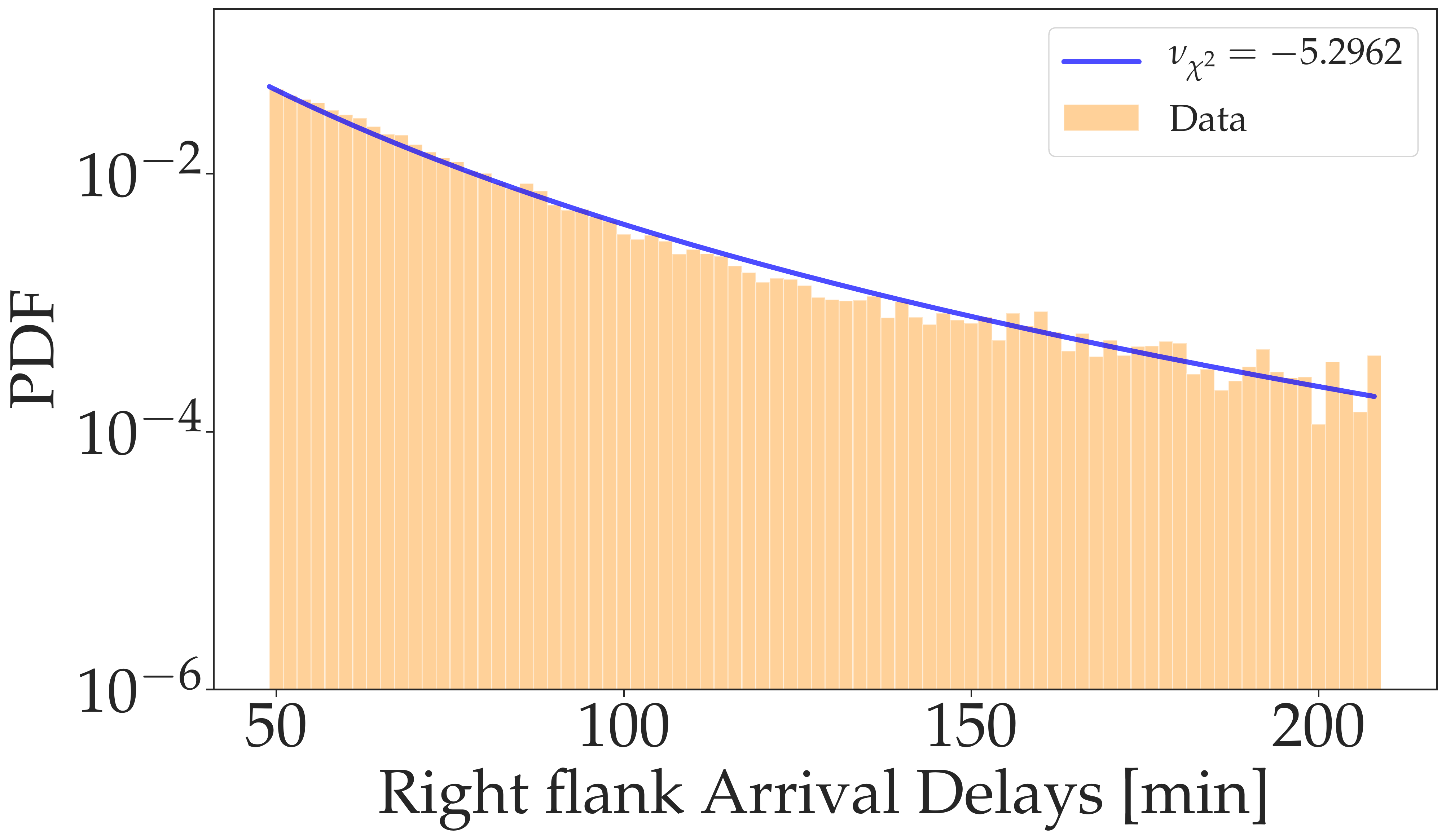}%
    \caption{Applying superstatistics leads to consistent results. Left: We extract the distribution of local exponents and compare them to a $\chi^2$ and inverse $\chi^2$ fit (based on the method of least squares).
    Right: Using the previously derived $\chi^2$ distribution, we again derive a $q$-exponential with right exponent $\nu_{\chi^2}\approx -5.296$, compared to the fitted one of $\nu_{\text{MLE}}\approx -5.371$. \revise{We note that the power-law decay of the data is well captured by the $q$-exponential induced by the $\chi^2$-distribution. The blue curve is scaled to the same amplitude as the data for visual guidance.}
    }
    \label{fig:Super_consistency}
\end{figure}

%Superstatistics consistency
Next, let us carry out an important consistency check: As explained above, the mixing of numerous local exponential distributions with exponents following a $\chi^2$-distribution leads to a $q$-exponential. Now, we can make a histogram of the $\lambda$-distribution and fit it with a $\chi^2$- and an inverse $\chi^2$-distribution. Then, we derive the $q$-exponential from the fitted $\chi^2$-distribution and compare it with the direct fit of the $q$-exponential and the original data. This is illustrated in Fig.~\ref{fig:Super_consistency}.

We note that the empirical $\lambda$-distribution is slightly better fitted by an inverse $\chi^2$- than a $\chi^2$-distribution, as also observed in other application areas \cite{chen2008superstatistical,williams2020superstatistical}. Overall, the superstatistical description seems consistent, given the short time series of flight delays under consideration. The $q$-exponential derived from the $\chi^2$ tends to overestimate the PDF at low values, which is understandable as we also exclude them for the fitting of the $q$-exponential via MLE (see Methods). Still, the tail behavior of the $q$-exponential based on the $\chi^2$ matches the real data and the MLE fit nicely. This means the observed power laws of the right flanks are essentially explained by a suitable superstatistics which describes changes in the microvariables on a time scale of $T\approx 1.5$ hours. 

\section{Connecting the flanks}
\label{sec:connecting}
So far, we focused our attention on describing and fitting the tail aspects of the distribution, namely the left, approximately exponential, flank and the right, approximately $q$-exponential, flank. Both these functions combined overestimate the peak of the distribution and hence, we also included the mean delay as the final metric in our framework. Now, let us consider how the two tail distributions could be merged in one smooth-fitting function. 

First, we note that the so far mostly ignored central part of the delay distribution \revise{can} be approximated by a Gaussian distribution, based on the parabola shape in the log-scale plots. We use this insight to propose the following continuous fitting function 
\begin{align}
p(t) = \begin{cases} 
    A_e \exp{\left(-\lambda \sqrt{C+(t-t_\text{peak})^2}\right)}, t<t_\text{peak}
    \\
    A_q \exp_q{\left(-\lambda_q \sqrt{C+(t-t_\text{peak})^2}\right)}, t\geq t_\text{peak}
    \label{eq:piecewise}
   \end{cases}
\end{align}
with $\exp_q(t)=(2-q)\lambda_q \left[ 1+ (q-1) \lambda_q t \right]^{\frac{1}{1-q}}$ being the $q$-exponential function. Here, $A_e$ and $A_q$ are amplitudes, $C$ is a curvature parameter, describing  the approximately Gaussian part in the center,  $t_\text{peak}$ is the delay at the peak of the delay distribution, where we split into left and right flanks and $t$ is the delay value, \revise{see Methods for fitting details and code.}

\begin{figure}[!ht]
    \centering
    \includegraphics[width=14cm]{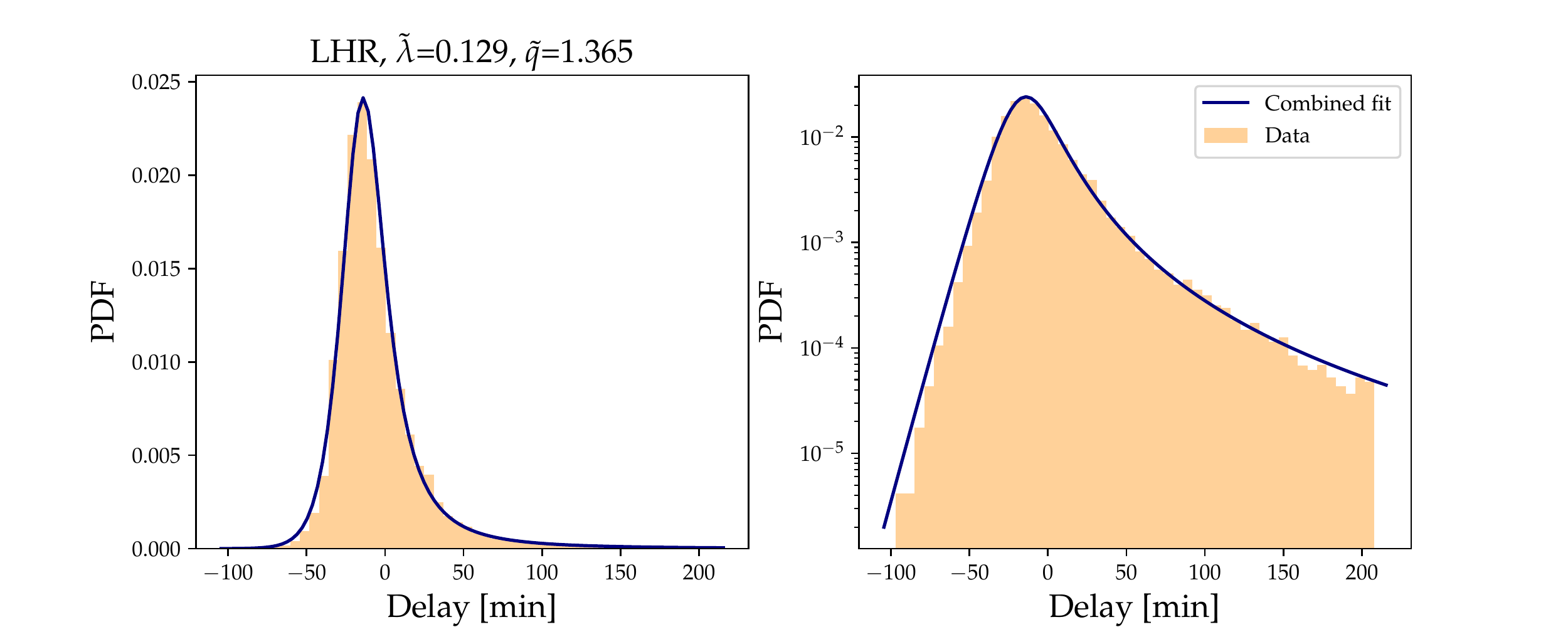}\\
    \includegraphics[width=14cm]{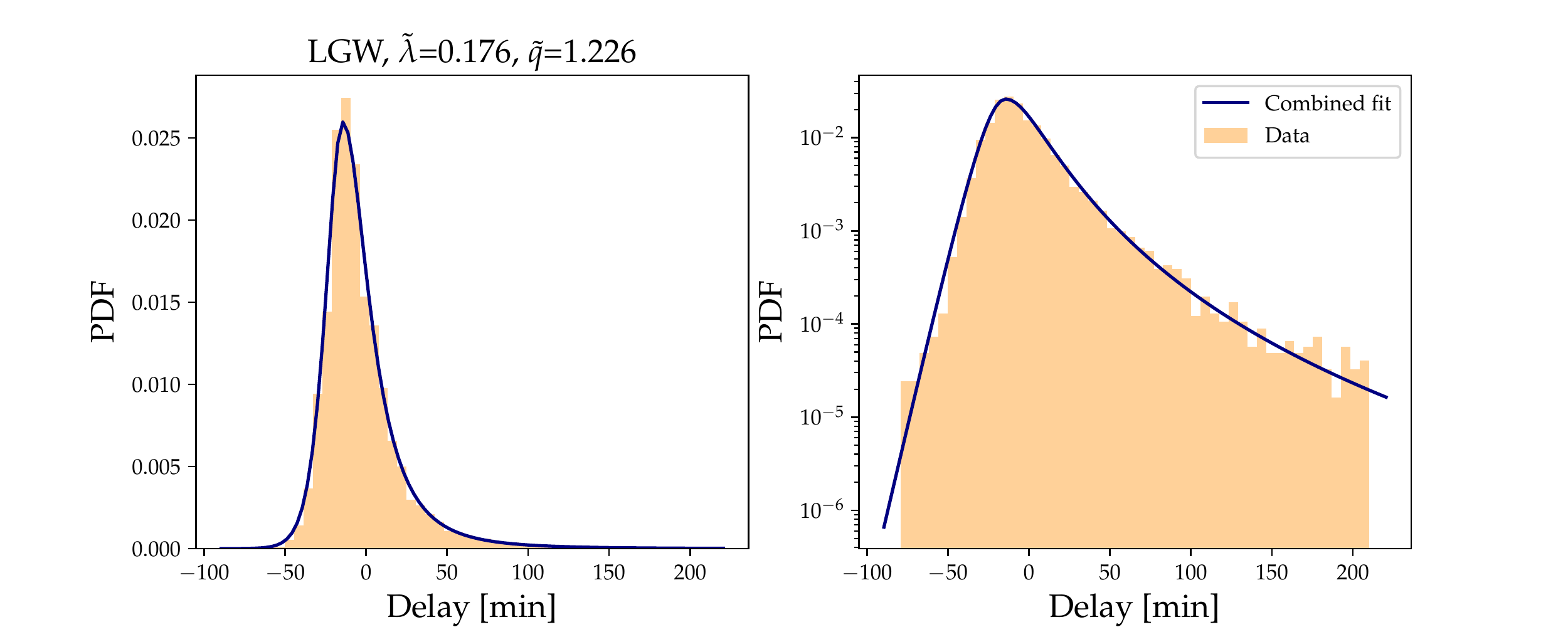}
    \caption{Using the approximately Gaussian shape in the center, we smoothly combine left and right flank fits into one coherent fit of the full delay data set. \revise{To emphasize the quality of the fit, we display both a linear (left) and logarithmic (right) scale of the PDF for LHR (top) and LGW (bottom), the two airports with the most flights in our data set.}}
    \label{fig:SmoothCombinedFit}
\end{figure}

The resulting fit is a smooth function, covering the full delay range, see  Fig.~\ref{fig:SmoothCombinedFit}. Since the new curvature parameter $C$ also influences the general shape, the new values for $q$ and $\lambda$, now named $\tilde{q}$ and $\tilde{\lambda}$, are slightly different from the ones solely focusing on the tails (empirically we tend to observe a slight reduction in $\lambda$ and increase in $q$).  Still, the general observations using the delay indices and comparing airlines, such as in Figs.\ \ref{fig:Compare_AirlinesInternational}-\ref{fig:Compare_Airports}, remain mostly unchanged. Equation \eqref{eq:piecewise} provides an alternative approach to the three delay indices introduced so far. If one is interested in describing the full distribution as accurately as possible, we recommend using equation \eqref{eq:piecewise}. Meanwhile, to compare performance of individual airlines or to obtain a general impression of the delay distribution, the three delay indices are a simplified framework, allowing easy and robust estimation and comparison. 
Finally, note that the full curve is not strictly a probability density function as we did not enforce that its integral equals one. While theoretically making it easier by reducing the number of parameters, that would make the fitting more difficult in practice as the integrals cannot be evaluated analytically by hand and impose additional constraints during the fitting. Also note that our observed flight delays are constrained to the finite interval $[-100,210]$, whereas the fitting function is defined on $[-\infty , \infty]$, which makes the normalization outside the interval ambiguous.

\section{Discussion and Conclusions}
%Summary
In summary, we have analysed a newly obtained data set of plane delays for various British airports, which contains tens of thousands of flights, aggregated over multiple months. \revise{We believe this is a substantial improvement on some earlier studies which, to the best of our knowledge, only investigated  a few days of measurements} and a couple of thousand flights, thereby greatly underestimating the contribution of the tails to the probability distribution \cite{caccavale2014model}. %(10 days, 7140 flights)
Interestingly, we find that all investigated airports and even individual airlines at each airport follow a qualitatively similar distribution, namely an approximately exponential decay on the left flank (of negative delays) and a slowly decaying power law on the right flank (of positive delays). 
To characterize these distributions and systematically compare airlines and airports, we have developed a framework to quantify delay performance. Critically, we do not merely use the mean delay but also consider extreme events of both positive and negative delays via their respective flanks in the empirical probability distribution.
Applying this newly developed framework, we find substantial differences between airlines serving short and long-distance routes. 

We offer an explanation for the emerging power law on the right flank via superstatistics: The local  $q$-exponential distribution with its heavy tails seems to arise from many superimposed exponential distributions. In particular, we identify the long time scale  $T$ as approximately 1.5 hours, during which delays fall off exponentially. Comparing to other superstatistical results \cite{beck-cohen, beck2001}, we note  
the relevance of both $\chi^2$-distributions and inverse-$\chi^2$-distributions for the scale parameter, similar to the ones observed in air pollution or cancer 
\cite{chen2008superstatistical,williams2020superstatistical}, stressing again the universality of superstatistics.
Finally, we propose a continuous function to capture the full delay statistics. While this introduces additional parameters and the superstatistical theory mentioned previously can no longer be used to rigorously derive the fitting function, this fit does describe the full distribution with high accuracy. 

%Discussion
Our framework of three delay indices to characterize flight delay distributions 
can be applied quite generally to
measure the punctuality of flights, going beyond an analysis based on just the mean. Crucially, while airlines or airports might be able to ``game" the system of mean delays, this is not possible with the left and right exponents. Companies could shift their flight schedule, i.e.\ announce intentionally that flights will take longer than they do in practice, and thereby systematically record early arrivals so pushing their mean delay to negative values. However, such a procedure would still leave the remaining two indices (left and right exponent) untouched so that they provide a stable way of measuring performance.

%Covid 19
One remarkable result is the impact of the global pandemic of COVID-19 on the delay statistics. Heathrow (LHR) under COVID-19 conditions (travel restrictions, quarantine upon arrival, etc)  displays an impressively low mean delay, while the left flank decay was mostly unchanged. Interestingly, LHR still experienced some relatively heavily delayed flights during the  COVID-19 pandemic, which leads to pronounced heavy tails towards the right and thereby a poor performance in the right exponent. 
\revise{These observations indicate that in different (COVID-19) situations and given fewer flights, airports can perform better in some aspects (e.g. mean delay) than under business-as-usual conditions, while other observables (extreme delays) can still  be improved.}
Aside from the upsides of COVID-19-related lockdown measures on air quality \cite{shrestha2020lockdown,schafer2020covid} or $CO_2$ emissions \cite{le2020temporary}, we find that having fewer flights also improves delay statistics. 

%assumptions 
We have assumed throughout this article that negative delays are preferred by all passengers. However, some passengers might value arrival at exactly the predicted time more highly than arriving early. This would change the interpretation of the left index slightly:  Instead of desiring low exponents, airlines and airports should aim for high exponents. Similarly, the absolute value of the delay should be zero, i.e.\ arrival on time should be the default. Regardless of preference, the indices, as introduced,  provide a sufficient framework to measure the delay performance.

%Outlook
In the future, we would like to apply our framework to delay statistics at other airports in different countries, and investigate how delays are related to geographical distance of the flights. In particular it would be interesting to see how our three indices differ between years, countries and so on.  From a more fundamental perspective, we aim to further understand correlations in the flight delays.  Preliminary indications from the British data are that on ``typical" days correlations decay quickly but on some ``exceptional" days (perhaps those where external factors affect many flights) the autocorrelation function can settle on a non-zero value for some time and many flights have long delays which contribute to the tail of the probability density function. Long-range  temporal  correlations and memory effects have been studied in many other physical and non-physical systems~\cite{Rangarajan03,Beran13}; modelling such effects here is challenging, since the build-up of delays at one airport may be influenced by earlier flights to and from completely different airports, but practically important since controlling the ``cascading" of delays would lead to a significantly improved passenger experience. \revise{In this way, future investigations could take into account spatio-temporal information from the entire worldwide air transportation network. More concretely, our data set could be expanded in type of information as well as volume. First, it would be interesting to also study departure delays, in addition to the arrival delays studied here. Furthermore, we could explicitly include flight duration and distance and investigate correlations between delays and flight distance/duration for many different airports in the world.}

%\begin{acknowledgments}
\subsection*{Acknowledgments}
This project has received funding from the European Union’s Horizon 2020 research and innovation programme under the Marie Sklodowska-Curie grant agreement No 840825.
%\end{acknowledgments}

\subsection*{Author contributions}
E.M., B.S., contributed equally. E.M., B.S., and C.B. conceived and designed the research. E.M. collected the data, E.M. and B.S. analysed the data and produced the figures. R.J.H. and all other authors contributed to discussing and interpreting the results and writing the manuscript. 

\subsection*{Competing interests}
The authors declare no competing interests.

\section*{Methods}
\subsection*{Data processing}
As we mentioned in the main text, \revise{for each flight, we recorded the airline company operating the flight, the flight number, the departure and arrival airports as well as the scheduled and actual landing times, as provided on the airport web page.} The data was cleaned and organized according to the delay, computed as the difference between scheduled arrival time and actual arrival time for each flight. We kept data for each arrival airport as well as a summary of the overall delays, independent of the arrival airport. A ``negative'' delay occurs when the actual aircraft arrival is earlier than the expected one, according to the scheduled timetable. After examining the data it became evident that a reasonable cut-off point as to how early or late an aircraft can arrive at the designated airport should be implemented. This prevents over-representation of individual extreme events in the resulting probability distributions. We decided that the delays (in minutes) would have to be contained in the interval $[-100, 210]$.

\subsection*{Theoretical distribution fitting}
Here we explain the fitting procedure in more detail. We approximate the empirical distribution of the left flank, where negative delays are dominant, with an exponential distribution of the form 
\begin{equation}
\label{exp1}
p(t_L;\lambda) = \lambda e^{- \lambda t_L}, \lambda >0.
\end{equation}
As we have seen in the main text, the observed distribution curves towards a Gaussian distribution around the peak value and thereby deviates from an exponential distribution. Hence, we restrict our fitting to values deviating from the central area as follows. Let $t_\text{peak}$ be the delay at which the distribution reaches its highest PDF value and $t_\text{min}$ the smallest delay we observe. Then, we restrict our exponential fit to any delay falling in the interval $[t_\text{min}, t_\text{peak}-0.3 |t_\text{min}-t_\text{peak}|]$, where $|...|$ indicates the absolute value. Following this restriction, we define the left flank delay values as \begin{equation}
    t_L=-t+t_\text{peak}-0.3 |t_\text{min}-t_\text{peak}|, t\in [t_\text{min}, t_\text{peak}-0.3 |t_\text{min}-t_\text{peak}|].
\end{equation}

We now turn to the right flank of the empirical distribution, i.e.\ the portion of the data set that constitutes the majority of the positive delays. The $q$-exponential is much better at incorporating parts of the Gaussian central distribution on the right-hand side than the exponential distribution is on the left flank. Hence, we only exclude the smallest $10\%$ of the data, i.e.\ we consider delays $t$ in the interval interval $[t_\text{peak}+0.1 | t_\text{max}-t_\text{peak}|,t_\text{max}]$, where $t_\text{max}$ is the highest delay observed. Hence the right-flank delays to be fitted are defined as
\begin{equation}
    t_R=t-t_\text{peak}-0.1 | t_\text{max}-t_\text{peak}|, t\in \left[t_\text{peak}+0.1 | t_\text{max}-t_\text{peak}|,t_\text{max}\right].
\end{equation}
Our theoretical distribution choice is now a $q$-exponential
\begin{equation}
\label{qexp}
 p (t_R; q, \lambda_q) = (2-q)  \lambda_q \left[ 1+ (q-1) \lambda_q t_R \right]^{\frac{1}{1-q}},
\end{equation} with parameters $ \lambda_q$ and $q$.
\revise{It has been shown that $q$-exponentials and $q$-Gaussians arise from maximizing Tsallis entropy \cite{tsallis1988possible}.}

Note that both $t_L$ and $t_R$ are defined such that they start at 0 and continue towards positive values to keep the fitting functions easier. 

These two functions (exponential and $q$-exponential) are fitted to the data using a maximum likelihood estimate (MLE), i.e.\ maximizing the Likelihood $L(\mathbf{\theta}, \mathbf{x})$. Here, $\mathbf{x}$ indicates the data we wish to fit and $\mathbf{\theta}$ the set of parameters that are being optimized. The likelihood of a parameter setting $\mathbf{\theta}$ on a given one-dimensional data set $\mathbf{x}=\left(x_1, x_2, ..., x_N \right)$ is computed as 
\begin{equation}
    L(\mathbf{\theta} , \mathbf{x})=\prod_{i=1}^N p( x_i,\mathbf{\theta}),
\end{equation} 
with probability density function $p( x_i,\mathbf{\theta})$, dependent on the parameters $\mathbf{\theta}$. Technically, we carry out the MLE using the \emph{scipy.stats} module in python with custom PDFs, see also Code availability (below) for a link to the code.

\subsection*{Fitting the smooth combined function}
To obtain a smooth fit, combining both flanks, we employ the following procedure. We first estimate the exponential decay rate $\lambda$ based on the lowest 70\% of negative delays, then estimate $q$ and the $q$-exponential decay rate $\lambda_q$ based on almost the full right-hand side of the histogram. This is identical to the procedure for the individual flanking fits. Next, we estimate the central curvature $C$, which we assume to be identical for both intervals, and the amplitudes $A_e$ and $A_q$, as well as $\lambda_q$ using least squares fitting. While carrying out this least-square fit, we also allow the parameters $q$ and $\lambda$ to vary  slightly from the MLE-optimal value determined earlier, while all other parameters are not bounded. The reason to allow any variance is to ensure a continuous fit while keeping the change from the optimal MLE parameters small. \revise{Empirically, we find that restricting $0.95 \ q_\text{MLE}\leq \tilde{q}\leq1.15 \ q_\text{MLE}$ and $0.95 \  \lambda_\text{MLE}\leq \tilde{\lambda}\leq1.05 \ \lambda_\text{MLE}$ yields the best results.}
Technically, we use the \emph{scipy.stats} module to perform the MLE fits and the least-square fit; continuity is ensured using constraints in the \emph{symfit} package.

\subsection*{Airline data}
In Figs.\ \ref{fig:Compare_AirlinesInternational} and \ref{fig:Compare_AirlinesLow} we compared several airlines. Let us briefly list how many flights we analysed to derive our delay indices:
For the short-distance airlines
``Wizz Air'': 2428, ``easyJet'': 15449, ``Ryanair'': 13488, ``Vueling'': 1034, ``Jet2'': 1215; for the other airlines we have 
``Iberia'': 12892, ``British Airways'': 38257, ``Aer Lingus'': 7331, ``Finnair'': 8560, ``American Airlines'': 23119, ``Air Canada'': 7247, ``United Airlines'': 6797, ``Japan Airlines'': 5966, ``Qatar Airways'': 5935.
For all airlines we have at least 1000 flights and often several thousand flights.

\subsection*{Data availability}
The original data of airport arrivals has been uploaded to an open repository:
\url{https://osf.io/snav9}.
All data that support the results presented in the figures of this study are available from the authors upon reasonable request.

\subsection*{Code availability}
Python code to reproduce figures,  perform the fits and extract the delay indices, is also uploaded here: \url{https://osf.io/snav9/}.

\bibliography{sample}

\end{document}